\documentclass[12pt]{article}
\usepackage{bm}
\usepackage{amssymb}
\usepackage{amsfonts}
\usepackage{graphicx}
\usepackage{caption}
\usepackage{subcaption}
\usepackage{soul}
\usepackage{hyperref}
\usepackage{geometry}
\usepackage{setspace}
\usepackage{authblk}
\title{\textbf{Effects of taxes, redistribution actions and fiscal evasion on wealth inequality: an agent-based model approach}}

\author[1,2,*]{Iago N. Barros}
\author[1,2,3]{Marcelo L. Martins}

\affil[1]{Departamento de Física, Universidade Federal de Viçosa, 36570-900, Viçosa, MG, Brasil}
\affil[2]{Ibitipoca Institute of Physics (IbitiPhys), Conceição do Ibitipoca, 36140-000, MG, Brasil}
\affil[3]{National Institute of Science and Technology for Complex Systems, CBPF, 22290-180, Rio de Janeiro, RJ, Brasil}
\affil[*]{\texttt{iago.barros@ufv.br}}

\begin{document}

\maketitle

\begin{abstract}
	In capitalist societies, only a single right can be fully exerted without constraints of any kind: the limitless accumulation of wealth. Such imperative or prime axiom is the ultimate cause of the raising waves of inequalities observed today. In this work we extended the agent-based model proposed by Castro de Oliveira \cite{Oliveira} to study the effects of non-uniform income redistribution policies and tax evasion on the final steady-state wealth distribution of economic agents. Our simulational results strongly support that well designed policies of income redistribution and rigid control of tax planning possibilities are unavoidable instruments to promote the raise of more economically egalitarian and sustainable societies.
\end{abstract}

\section{Introduction}

Currently, economic inequality and redistribution issues laid at the core of political struggle. The scenario is so dramatic that, for instance, according the Global Wealth Report 2023, produced by the UBS Bank \cite{UBS}, almost half ($48.4 \%$) of the Brazilian wealth is concentrated on the hands of its $1 \%$ richest, while $58,7 \%$ of its population suffer from food insecurity at some level.  Comprising the top five of inequality rank are India ($41 \%$), USA ($34 \%$), China ($31.1 \%$), and Germany ($30 \%$). Early, the United Nations Development Program (UNDP) in its 2019 Human Development Report \cite{UNDP}, raised a vigorous alert: ``inequality is framed around economics, fed and measured by the notion that making money is the most important thing in life. But societies are creaking under the strain of this assumption, and while people may protest to keep pennies in their pockets, power is the protagonist of this story: the power of the few; the powerlessness of many; and collective power of the people to demand change''. Given the severity of such situation, a certain consensus concerning a few basic principles of social justice has emerged: a fair society must maximize opportunities and lowest life standards provided by the social system --- the ``maximin'' directive \cite{Kolm,Rawls}. However, it is the failure to address these systemic challenges that will further entrench inequalities and consolidate the power and political dominance of the few.

Several causes fuel wealth to concentrate. In USA and United Kingdom, the major component is the increasing wages' inequality, strongly reinforced by decreasing tax progressivity and social transferences since 1970s. In EU and developing countries, the primary source of inequality seems to be unemployment and underemployment \cite{Piketty}. The evolution of wealth distribution from $1987$ to $2013$, in which the greatest world fortunes increased at a rate of $6-7 \%$ per year against an average income growth rate of $1.5-2 \%$ \cite{Piketty}, suggests progressive taxes on capital as the unique way, under the capitalist mode of production, to control this inegalitarian dynamics. However, progressive taxation is currently under severe academic and political threats because its distinct functions remains insufficiently debated and varied categories of income are gaining exemption from tax law.

Despite have developed basically all materials, means of production, and technologies employed along the last two centuries, physicists barely contributed to criticize the diverse social orders established in this period. Meanwhile, a plethora of public policies needy of minimal rational or empirical foundations were implemented based on either faith or shallow ideologies. However, fast advances on modeling, analysis and simulations of complex systems are currently opening the road to physicists propose quantitative mathematical models that can reintroduce rationality into debate about diverse possibilities of social engineering. Then, concerning economic inequality and redistribution issues, Bouchaud and M\'ezard \cite{Bouchaud} and Castro de Oliveira \cite{Oliveira,Oliveira2} unveiled dynamically induced transitions between an economy dominated by a single or few individuals and a regime in which the total wealth remains distributed among many agents. Moreover, Cardoso, Gon\c{c}alves, and  Iglesias \cite{Ben-Hur} indicated that economic inequalities increase during periods without social protection and decrease when such policies are restored. Calvelli and Curado \cite{Calvelli} were able to qualitatively reproduce real wealth distributions and show that distinct taxation procedures can not prevent economy evolves into stationary states of extreme wealth concentration. Regarding wealth distributions, distinct functional forms, namely, exponential \cite{Drago}, log-normal \cite{Souma}, gamma \cite{Lux,Silver}, Gaussian \cite{Iglesias}, power law \cite{Bouchaud,Calvelli,Drago,Chakrabarti} and Tsallis \cite{Soares} distributions were found for different countries and simple models. To sum up, power law behavior for the wealthiest $5$ or $10\%$ of the population, and gamma or log-normal distribution for the rest result from such limited empirical studies \cite{Chakrabarti}. Unfortunately, these developments still represent only a marginal approach.

In this paper the agent-based model proposed by Castro de Oliveira \cite{Oliveira} to study the final steady-state distribution of wealth generated by an annual random multiplicative process subjected to a linear wealth-dependent taxation rate was extended. Specifically, non-uniform income redistribution policies and tax evasion strategies (fiscal engineering) were introduced in order to understand their effects on wealth distribution and economic inequality dynamics.

\section{Modeling wealth dynamics}
\label{sec:model}
The model assumes that economic agents operate freely and independently as well as income redistribution occurs only through taxes and fiscal transferences, the so called pure redistribution regime. Accordingly, the wealth distribution in a society comprised by $N$ agents evolves by iteracting sincronously at discrete time steps the following rules proposed by Castro de Oliveira \cite{Oliveira}:

\noindent 1. The wealth $W_i$ of each individual $i$ ($i=1,2,\ldots,N$) is multiplied by a random factor $f_i$ chosen from a given probability distribution, such that $W_i$ changes to $W_i^\prime =f_i W_i$;

\noindent 2. Then, every agent $i$ is taxed based on their wealth $W_i$, the paid tax being $(A + p\omega_i) W_i$. Here, $A$ and $p$ are the consumption duty and the income tax, respectively. Also, $\omega_i=W_i/S$ is the fraction of total wealth $S = \sum_j W_j$ held by individual i (its wealth share). The model parameters $A$ and $p$ satisfy the constraints $0 < A < 1$ and $0 < A + p < 1$. After taxation, the wealth of agent $i$ changes from $W_i^\prime$ to $W_i^{\prime \prime}=(1-A-p \omega_i) W_i^\prime$.

\noindent 3. Finally, a fraction $R$ of the total collected taxes, $\sum_n (A+p\omega_n)W_n$, is redistributed among all agents, eventually increasing their wealths to $\{W_n^{\prime \prime \prime}\}$. An uniform redistribution among all agents, leading to $W_n^{\prime \prime \prime}=W_n^{\prime \prime}+R \sum_i (A+p\omega_i)W_i^\prime/N$, was considered in reference \cite{Oliveira}.

After applying these three evolution rules to all agents, a new time step (``year'') starts from the set $\{W_n^{\prime \prime \prime}\}$, and so on. As it was originally implemented \cite{Oliveira}, the multiplicative factors $f_n$ were randomly selected with equal probabilities to be either $f_n = 2$, which double the current wealth $W_n$, or $f_n = 1$, leaving the wealth $W_n$ unchanged. In turn, different values for parameters $A$, $p$, and $R$ (including distinct functional forms for the redistribution $R=R(W_n)$) were tested in our computer simulations of an extended Castro de Oliveira model. We focused on the time evolution of wealth distributions as well as on the dynamics of $\omega_{max}$, the greatest wealth share owned by an individual. Indeed, as reported by Castro de Oliveira \cite{Oliveira}, $1-\omega_{max}$ is the order parameter for a phase transition between an absorbing state $(\omega_1=1, \omega_n=0 \,\,\, \forall n\ge 2)$, in which all the wealth is owned by a single agent and the collapsed economy ceases to evolve, and an active phase in which the economy evolves forever with the wealth distributed
among agents. Such transition occurs for $R=0$ (in the absence of redistribution) at $p=0$. For $p<0$, corresponding to a regressive regime (the poor pay higher tax rates than the
riches), the collapsed phase is observed. In contrast, for $p>0$, the progressive regime (the riches pay higher tax rates than the poor), the active phase occurs. Furthermore, a redistribution $R \neq 0$ plays the role of an external field that destroys this transition. 

\section{Results}
\label{sec:result}
 \subsection{The model as a mapping}
The three evolution rules for wealth dynamics aforementioned can be translated into a discrete time, $N$-dimensional mapping:

\begin{eqnarray}
\label{map}
W_i(t+1) & = &\left[1-A-p \frac{f_i(t)W_i(t)}{\sum_jf_j(t)W_j(t)} \right]f_i(t)W_i(t) \nonumber \\ 
         &   & + \frac{R}{N} \sum_i \left[ A+p \frac{f_i(t)W_i(t)}{\sum_jf_j(t)W_j(t)} \right]f_i(t)W_i(t), 
\end{eqnarray}
where an uniform redistribution protocol was assumed. Through the analytical investigation of the map's fixed points and their stabilities, we can obtain some simple but interesting results.

Firstly, if $R=0$ -- a non-welfare fiscal state, the mapping (\ref{map}) become 

\begin{equation}
    W_i(t+1) = \left[1-A-p \frac{f_i(t)W_i(t)}{\sum_jf_j(t)W_j(t)}\right]f_i(t)W_i(t).
    \label{mapR0}
\end{equation}


\noindent This map, due to its monotonic properties—demonstrated in the stability analysis—has only one fixed point: the trivial one, \(W_i(t) = 0\) for all \(i\). This becomes evident when both \(A = 0\) and \(p = 0\), representing a laissez-faire economy, where the map decouples in independent stochastic binomial multiplicative processes \(W_i(t+1) = f_i(t)W_i(t) \) and the unique fixed point corresponds to a “dead” economy. When considering income taxes (\(p \ne 0\)), this fixed point \(\mathbf{W_0}\) becomes singular, indicating that the system is non-analytical at this point. Nevertheless, it still behaves as an attractor or repeller.

In terms of wealth share $\omega_i$, mapping (\ref{mapR0}) has two distinct classes of non-trivial fixed points. One of them, $w_i=1/N^\prime$ for $i \le N^\prime$ and $w_i=0$ for $i \ge N^\prime$ ($N^\prime \le N$) occurs if and only if $f_i(t)=f(t)$ $\forall t$ and $i \le N^\prime$, i.e, the wealth of all active agents (with $\omega\ne0$) evolve under the same sequence of random multiplicative factors. In particular, the case $N^\prime=N$ is very emblematic. It corresponds to a utopian, full equality society in which all agents have both alike economic skills to enlarge their assets and the same initial wealth share. Clearly, this ``naive communist'' or homogeneous fixed point is unstable. This is also true for $2 \le N^\prime <N$. The other class of non-trivial fixed points is $w_i=1$ and $w_j=0$ $\forall j \ne i$ and $i \in [1,N]$, that is $N$-fold degenerate. All such $N$ fixed points correspond to a collapsed economy in which only a single agent plays the game and owns the total wealth. According to Castro de Oliveira simulations \cite{Oliveira}, these fixed points $\mathbf{w}_c$ are stable for regressive taxes but unstable for progressive ones.

In addition to these two classes, there are random aperiodic trajectories in $N$-dimensional phase space $\{\omega_1,\omega_2,\ldots,\omega_n\}$ having certain statistical features that define dynamical stationary states. Specifically, in such trajectories the order parameter $1-\omega_{max}$ does not vanishes \cite{Oliveira}. 

In his work Castro de Oliveira, primarily worried with the stability of the fixed point $\omega_{max}=1$, did not discuss the dynamics of wealth itself and hence the stability of the trivial fixed point $\mathbf{W_0}$. Thus, an imperative issue to be addressed is determine what regions in model parameter space ($A,p,R$) are controlled by this emblematic fixed point. 

In order to answer this question, we introduce two auxiliary maps. The first one is


\begin{equation}
V_i(t+1)=(1-A-p)f_i(t) V_i(t),
\label{bmp}
\end{equation}
corresponding to independent wealth multiplicative processes under taxation. The second one is 

\begin{equation}
    X_i(t+1) = (1-A)f_i(t) X_i(t),
\label{cmp}
\end{equation}
which is simply the case of taxation reduced to consumption tax alone.

For $p \ge 0$ (non regressive taxes) and $R=0$, the inequality

\begin{equation}
V_i(t) \le W_i(t) \le X_i(t),
\label{ineq}
\end{equation}
is valid, as one can easily see. Furthermore, these mappings, eqs. (\ref{mapR0}), (\ref{bmp}) and (\ref{cmp}), share the trivial fixed point. Therefore, in consequence of inequality (\ref{ineq}), where in parameter subspace ($A,p>0,R=0$) the trivial fixed point is unstable for mapping $V$, it will be for maps $W$ and $X$ too.

A perturbation $\boldsymbol{\epsilon}_0=(\epsilon_1,\epsilon_2,\ldots,\epsilon_N)$ around $\mathbf{W_0}$ under the mapping $V$ generates the orbit

\[V_i(t)=(1-A-p)^t \epsilon_i \prod_{j=1}^t f_i(j), \] 
after $t$ iterations. Averaging over all possible random sequences of binary multiplicative factors $f_i=z_1$, chosen with probability $q_1$, or $f_i=z_2$, selected with probability $q_2$, one obtains

\[ \langle \prod_{j=1}^t f_i(j) \rangle= \sum_{n=0}^{t} \left( 
\begin{array}{c} t \\ n \end{array} 
\right) q_1^n q_2^{t-n} z_1^n z_2^{t-n} =(q_1z_1+q_2z_2)^t. \]
Given that $q_1=q_2=1/2$, $z_1=2$ and $z_2=1$, this average is $(3/2)^t$. Thus,

\[ \langle V_i(t) \rangle =(1-A-p)^t \left( \frac{3}{2}\right)^t \epsilon_i = \epsilon_i e^{t \, ln \left[ \frac{3}{2}(1-A-p) \right] }. \]
Accordingly, $\langle V_i(t) \rangle \rightarrow \infty$ when $t \rightarrow \infty$ if $A+p<1/3$, and the trivial fixed point is unstable. Therefore, the inequality $A+p<1/3$ establishes a lower bound for the instability threshold associated to $\mathbf{W_0}$ on both maps $W$ and $X$. Given that $0<A<1$ and $0<A+p<1$ by model definition, the trivial fixed point of mapping (\ref{map}) is unstable within the triangle with vertexes at $(0,0)$, $(1/3,0)$, and $(0,1/3)$ in the parameter subspace $(A,p>0,R=0)$.

In turn, for $p<0$ (regressive taxes) inequality (\ref{ineq}) is no longer valid, but now

\begin{equation}
    \label{ineq2}
    V_i(t) \ge W_i(t) \ge X_i(t).
\end{equation}


\noindent The stability analysis of $\mathbf{W_0}$ under $W$ can be performed considering the orbit based on perturbation $\mathbf{\delta}_0=(\delta,0,\ldots,0)$. Since

\[ \begin{array}{lcl}
W_1(t) & = & (1-A+|p|)^t \, \delta \, \prod_{j=1}^{t} f_1(j) \\
W_i(t) & = & 0, \,\,\, i \neq 1,
\end{array} \]
one obtains  

\[ \langle W_1(t) \rangle = \delta e^{t \, ln \left[ \frac{3}{2}(1-A+|p|) \right] }. \] 
Accordingly, for $p<0$, $\langle W_1(t) \rangle \rightarrow \infty$ when $t \rightarrow \infty$ if $A-|p|<1/3$, and the trivial fixed point is unstable within the trapezium with vertexes at $(0,0)$, $(1/3,0)$, $(1,-2/3)$ and $(-1,-1)$. This results again impose lower bounds for instability borders in $W$ map.

Taken together, these results reveal that the trivial fixed point of $W$ is unstable within the region delimited by the straight lines $A=0$, $p=1/3-A$, $p=-A$ and $A=1$ of the parameter subspace $(A,p,R=0)$.


In addition, we can demonstrate that wealth dynamics is asymptotically driven towards the singular trivial fixed point for $p>A-\frac{1}{3}$ and $A>1/3$. Indeed, from (\ref{ineq}), if the trivial fixed point $\mathbf{W_0}$ is stable inside a region of parameters $(A,p)$ for $X_i(t)$ it will be stable for $W_i(t)$, once that $W_i(t) \ge 0$ for all $i$.

Analogous to what was previously made, it can be shown that the orbits generated by the mapping $X_i(t)$ are

\[X_i(t)=(1-A)^t \epsilon_i \prod_{j=1}^t f_i(j), \]
and, the average mapping of $\langle X_i(t) \rangle$ converges to the fixed point $\mathbf{W}_0$ for $p>0$ if $\frac{3}{2} (1-A) \le 1$, therefore for $A>1/3$. Furthermore, $X_i(t)$, and not only it's average, vanishes asymptotically for $A>1/2$, because $f_i(t) \leq 2$ and (\ref{cmp}) implies $X_i(t+1) \le X_i(t)$ $\forall$ $t$.

Similarly, for $p<0$ (regressive taxes) the inequality (\ref{ineq2}) tells us that if the stability of the fixed point is guaranteed for the mapping $V$ it will also be granted for $W$ and $X$. Since the orbit of $V_i(t)$ based on $\bm{\epsilon}$ for regressive taxes is given by

\[V_i(t)=(1- A + |p|)^t \epsilon_i \prod_{j=1}^t f_i(j), \]

\noindent the averaging over the random multiplicative factors yields that $\langle V_i(t) \rangle \to 0$ provided that $\frac{3}{2}(1-A+|p|) < 1$, thus for $A>\frac{1}{3} + |p|$. Moreover, if $A \ge \frac{1}{2} + |p|$ the mapping $W_i(t)$ decreases monotonically in time. Taking together, these results ensure the stability of the singular trivial fixed point within the region $p > \frac{1}{3} - A$ and $A>1/3$ of the parameter subspace $\{A, p, R=0\}$, as shown in Figure \ref{wealthDiagram}.

Our previous analysis is unable to provide insights into the stability of the trivial fixed point within the region defined by \(p > \frac{1}{3} - A\) and \(A < \frac{1}{3}\). To address this region and thus complete the stability diagram, we performed a mean-field approximation for the mapping $W$.

\subsection{Mean-field-like Approximation}

\quad\quad Since the tax paid by every agent $i$ is proportional to his wealth-share $\omega_i$ at time $t$, the mapping (\ref{mapR0}) can be written as

\begin{equation}
    W_i(t+1) = \left[1-A-p\omega_i(t)\right]f_i(t)W_i(t).
    \label{mapR0V2}
\end{equation}

Considering a uniform initial condition ($W_i(0) = W_0$ $\forall$ $i$) and $p>0$ (progressive taxes), all agents are "equivalent", in the sense that in a certain realization the agent $i$ will be the luckiest, but in another one he will be the poorest. Since there are $N$ agents and each possibility is equally probable, the long term mean wealth-share will be:

\[
\langle \bm{{\omega}}(t)\rangle = (1/N, 1/N, ..., 1/N). 
\]

Assuming that $\omega_i(t) \sim \langle \bm{\omega}(t)\rangle_i = 1/N$ the mapping becomes decoupled:

\[
W_{i}(t+1) = \left(1 - A - \frac{p}{N}\right) f_{i}(t) W_{i}(t).
\]
The orbit generated by this mapping from the state $(W_0, W_0, ..., W_0)$ after $t$ time-steps is

\begin{equation}
    W_i(t)=\left(1-A-\frac{p}{N}\right)^{t} W_0 \prod_{j=1}^t f_i(j). 
\end{equation}
Thus, for the progressive case, the trivial fixed point is stable if:

\[
\frac{3}{2} \left(1 - A - \frac{p }{N}\right) <  1 
\Rightarrow A > \frac{1}{3} - \frac{p}{N} \simeq \frac{1}{3}, \quad \mathrm{for\ } N \gg 1.
\]

\begin{figure}
    \centering
    \begin{subfigure}[b]{0.5\textwidth}
        \centering
        \includegraphics[width=\textwidth]{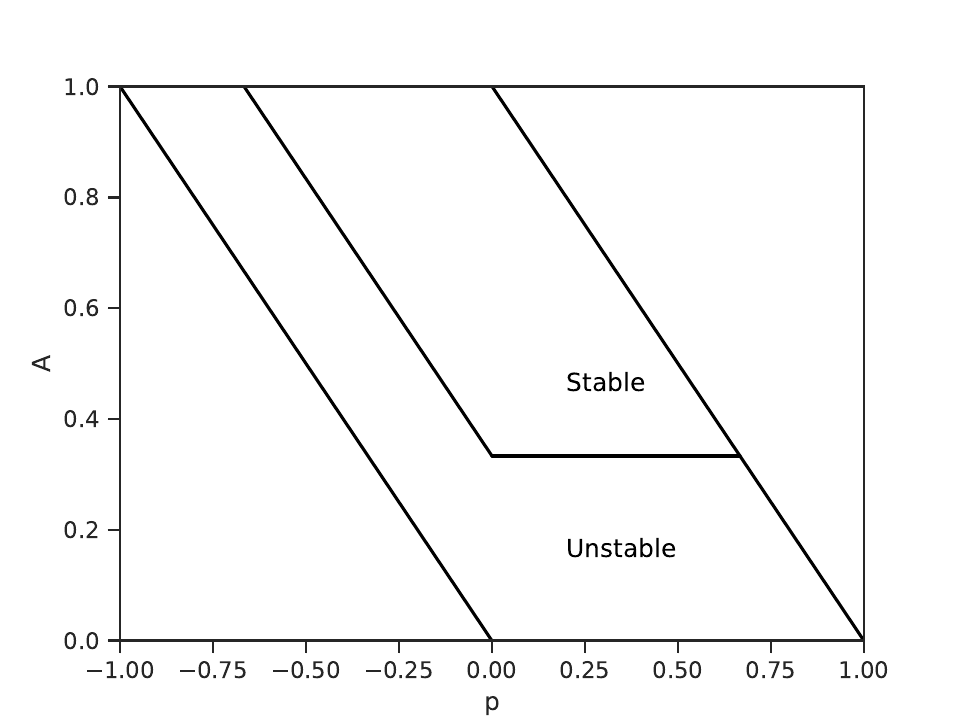}
        \caption{Stability of $\mathbf{W_0}$}
        \label{wealthDiagram}
    \end{subfigure}%
    \begin{subfigure}[b]{0.5\textwidth}
        \includegraphics[width=\textwidth]{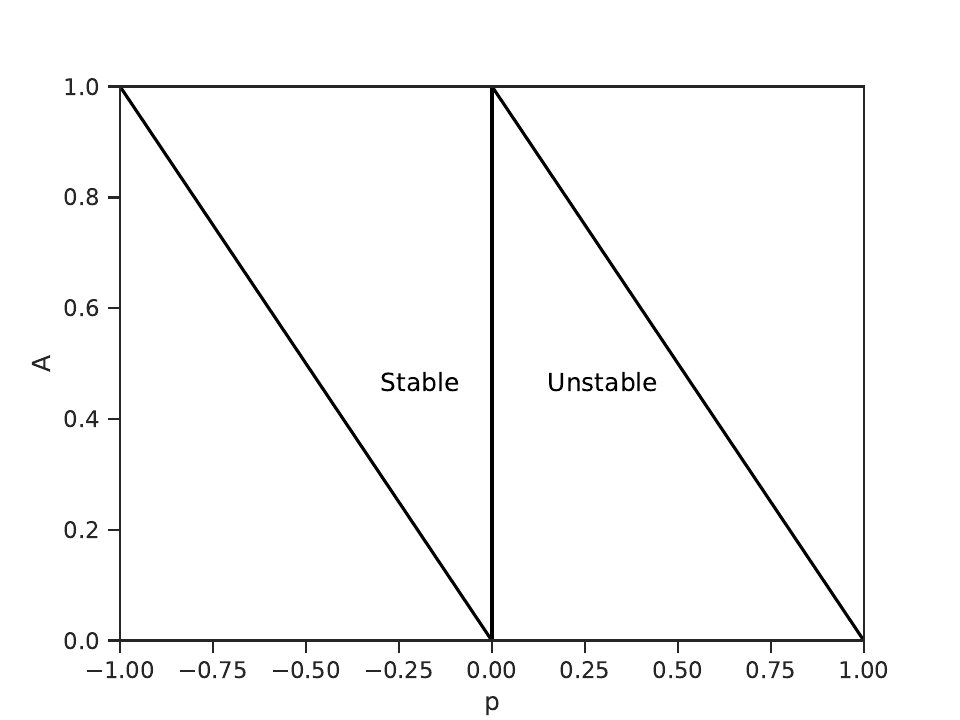}
        \caption{Stability of $\omega_{max}=1$}
        \label{wealthshareDiagram}
    \end{subfigure}

    \caption{Stability diagram for the model without redistribution. In panel (a), the filled lines represent the boundaries defined by \(A+p=0\), \(A+p=1/3\), and \(A+p=1\). The stable region for the trivial fixed point $\mathbf{W_0}$ was proven analytically, while the full unstable region was derived in the limit of large \(N\) using analytical methods and a mean-field-like approximation. In panel (b), we present the stability diagram for the fixed point \(\omega_{max}\) for comparison. For regressive taxes (\(p<0\)), the system asymptotically approaches the state \(\omega_{max}=1\). In contrast, under progressive taxes, the state \(\omega_{max}=1\) becomes an unstable fixed point.}

    \label{stabilityDiagrams}
\end{figure}

\noindent Similarly, it will be unstable if $A < \frac{1}{3} - \frac{p}{N} \simeq \frac{1}{3}$ for $N \gg 1$. And, with this, we have determined the stability of the fixed point $\mathbf{W_0}$ in the entire space of parameters $(A,p, R=0)$. The regions of stability for the fixed points $\mathbf{W_0}$ and $\omega_{max}=1$ are summarized in Figure \ref{stabilityDiagrams}.

\subsection{Wealth distribution for $p=0$}

In the absence of income tax (p=0) and redistribution (R=0) the mapping is given by 

\begin{equation}
    W_i(t+1) = [1-A] f_i(t) W_i(t).
    \label{p_0_mapping}
\end{equation}

\noindent Supposing an uniform initial condition $W_i(t=0)=W_0$ for all $i=1,2,...,N$, after t time-steps we get 

$$ W_i(t) = \prod_{k=0}^t f_i(k) [1-A]W_0 ,$$

$$ \log(W_i(t)) = \sum_{k=0}^t \log(f_i(k)) + \log([1-A]W_0) .$$

By the Central Limit Theorem, the distribution of the sum will converge to a Gaussian as $t\to \infty$. This means that wealth, in this case, will reach a lognormal distribution. For progressive and regressive taxes ($p\ne 0$), this analysis is more difficult to perform because we cannot ensure that the random variables are independent and identically distributed. As seen in Figure \ref{gauss}, wealth in these cases continues to present a lognormal distribution.

\subsection{Testing against computational simulations}
In order to further characterize the stable dynamical behaviors emergent from the Castro de Oliveira agent based model or, equivalently, mapping (\ref{map}), we simulated a system comprised by $N=1000$ agents, random  multiplicative factors $f=1$ or $f=2$, selected with equal probabilities at each time step by every agent, and $R=0$ (without redistribution) for distinct values of parameters $A$ and $p$.

\begin{figure}
    \centering
    \begin{subfigure}[b]{0.5\textwidth}
        \centering
        \includegraphics[width=\textwidth]{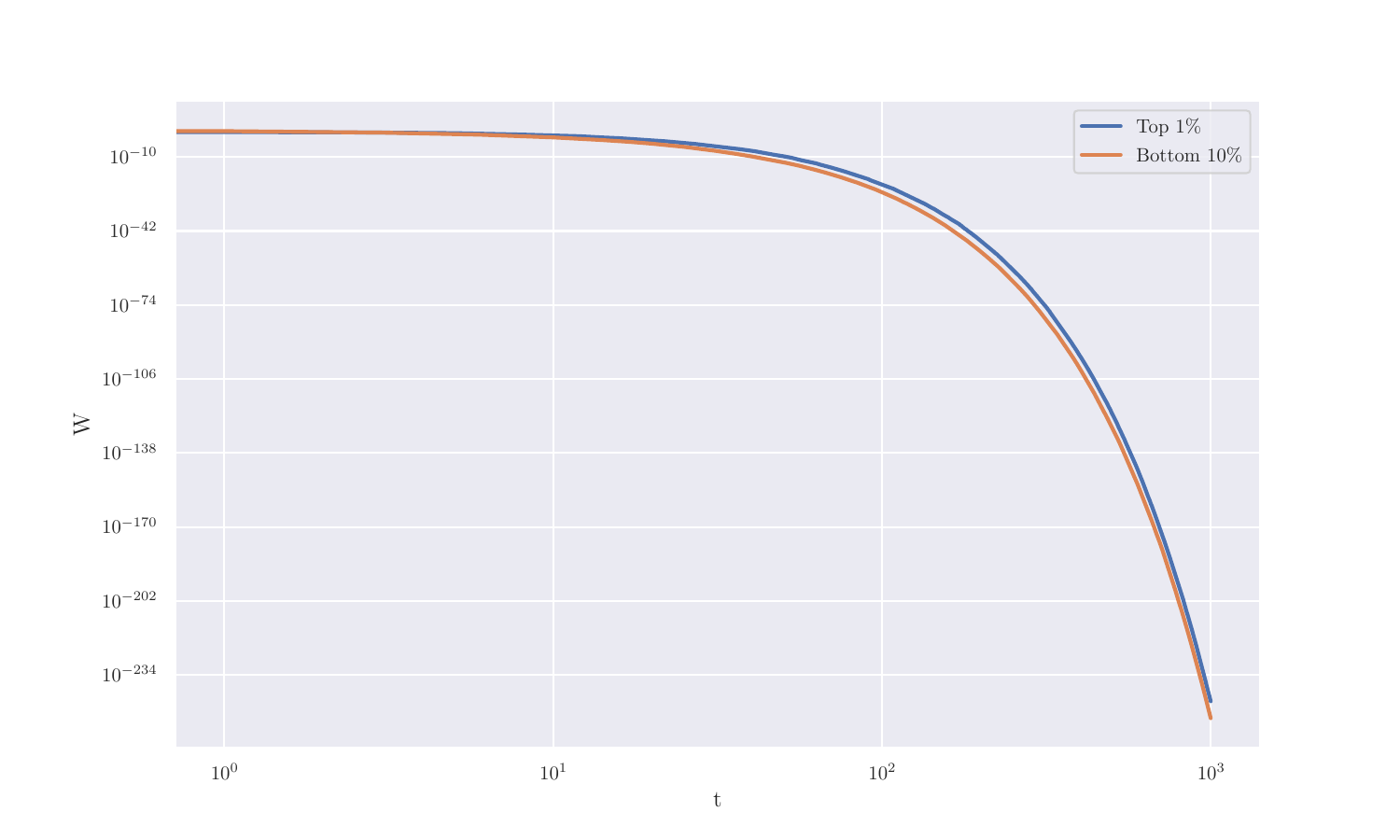}
        \caption{Impoverishment}
        \label{evolveWa}
    \end{subfigure}%
    \begin{subfigure}[b]{0.5\textwidth}
        \includegraphics[width=\textwidth]{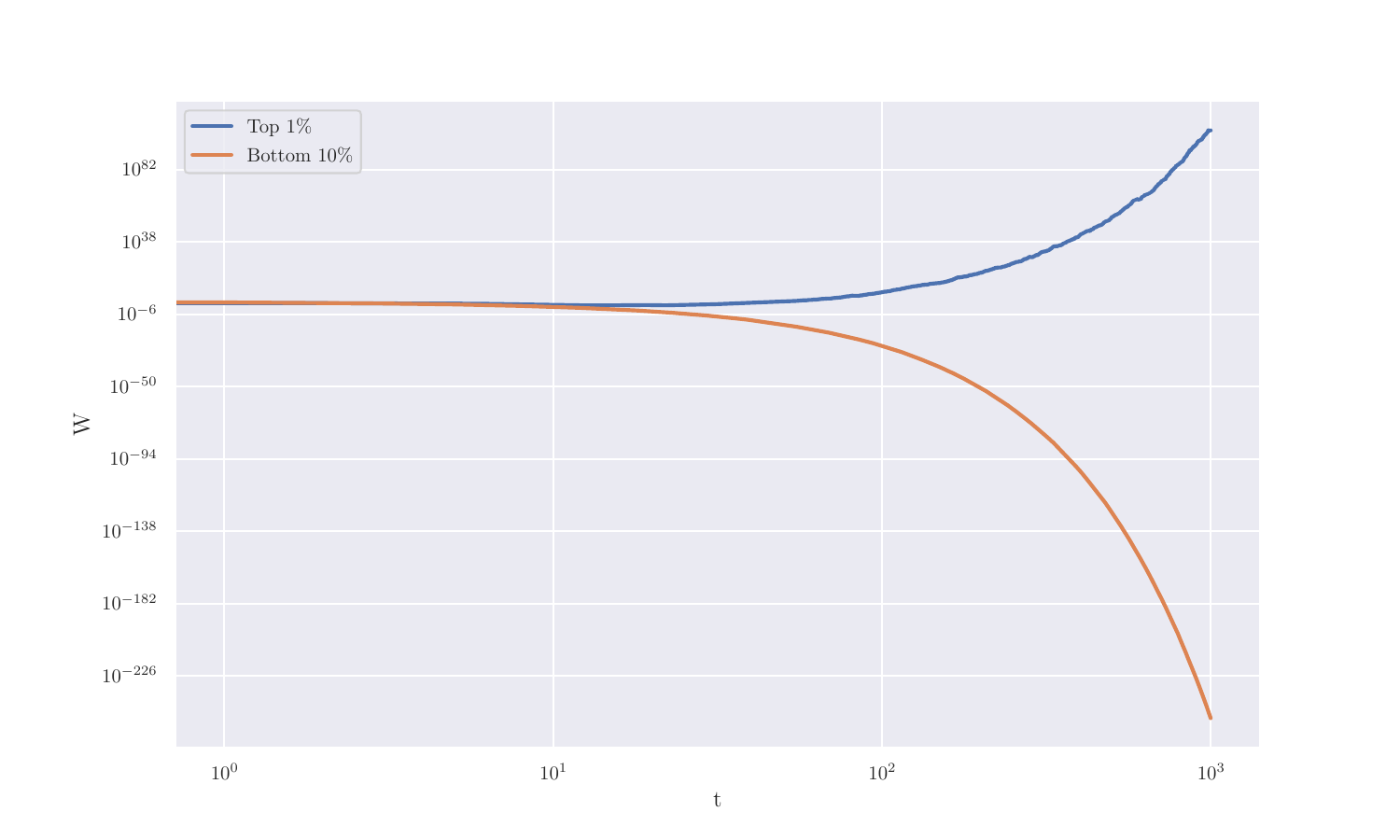}
        \caption{Enrichment and Dispossession}
        \label{evolveWb}
    \end{subfigure}
    \begin{subfigure}[b]{0.5\textwidth}
        \includegraphics[width=\textwidth]{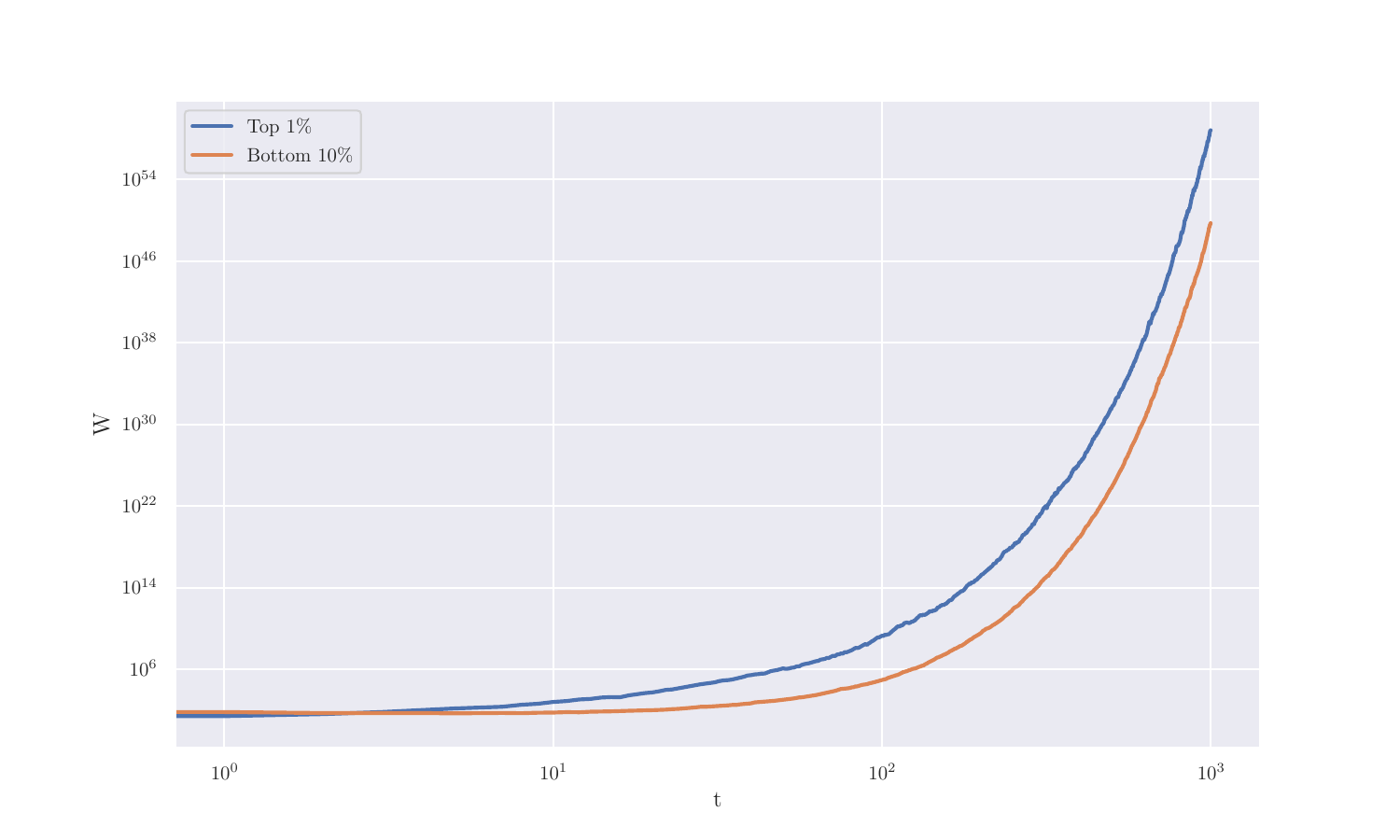}
        \caption{Global Prosperity}
        \label{evolveWc}
    \end{subfigure}

    \caption{Time evolution of total wealth owned by the 1\% richest (orange) and 10\% poorest agents (blue curve) without redistribution policies $R=0$.}
    \label{evolveW}
\end{figure}

In Figure \ref{evolveW} are shown three typical behaviors observed for the evolution in time of total wealths owned by the $1\%$ richest and $10 \%$ poorest agents. We called these dynamics (i) general impoverishment, in which all agents becomes progressively poorer, as illustrated in Figure \ref{evolveWa}; (ii) enrichment at the expense of dispossession, in which riches get even richer but the poor agents get poorer, as shown in Figure \ref{evolveWb}, and (iii) global prosperity, in which all agents become increasingly rich (see Figure \ref{evolveWc}).

\begin{figure}
    \centering
    \includegraphics[width=0.8\textwidth]{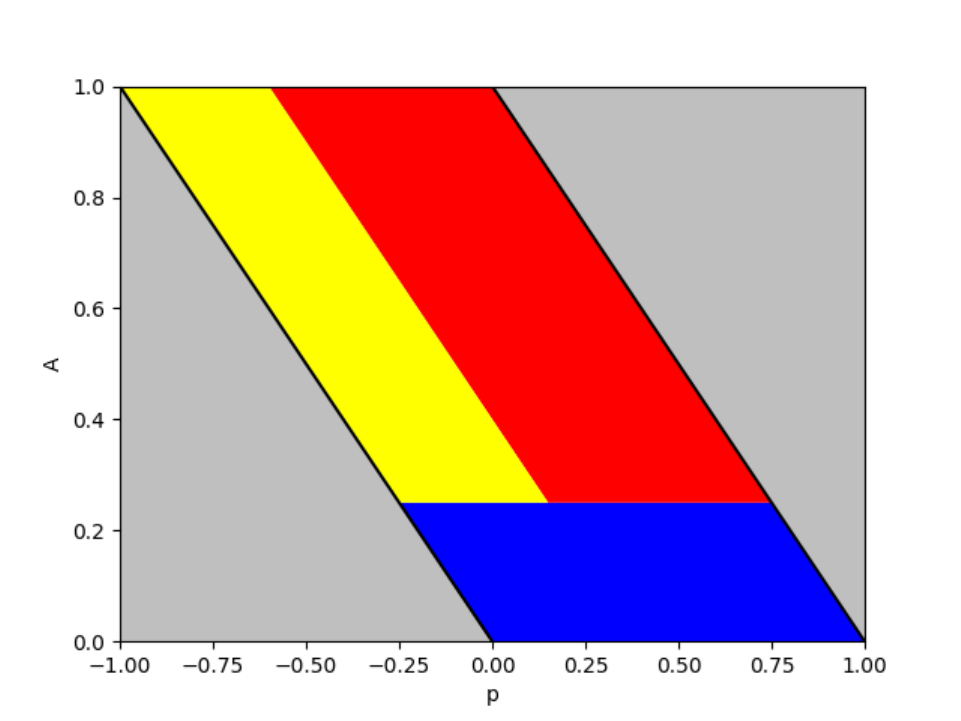}
	\caption{The regions of subspace parameter $R=0$ dominated by general impoverishment (red), enrichment at the expense of dispossession (yellow), and global prosperity (blue).}
	\label{parts}
\end{figure}

Our simulations indicate that such behaviors partition the parameter subspace $R=0$ in regions, shown in Figure \ref{parts}, qualitatively consistent with the analytical results obtained for the fixed point stability analysis of the wealth share mapping $\omega_i=W_i/\sum_nW_n$.

\begin{figure}
    \centering
    \begin{subfigure}[b]{0.5\textwidth}
        \includegraphics[width=\textwidth]{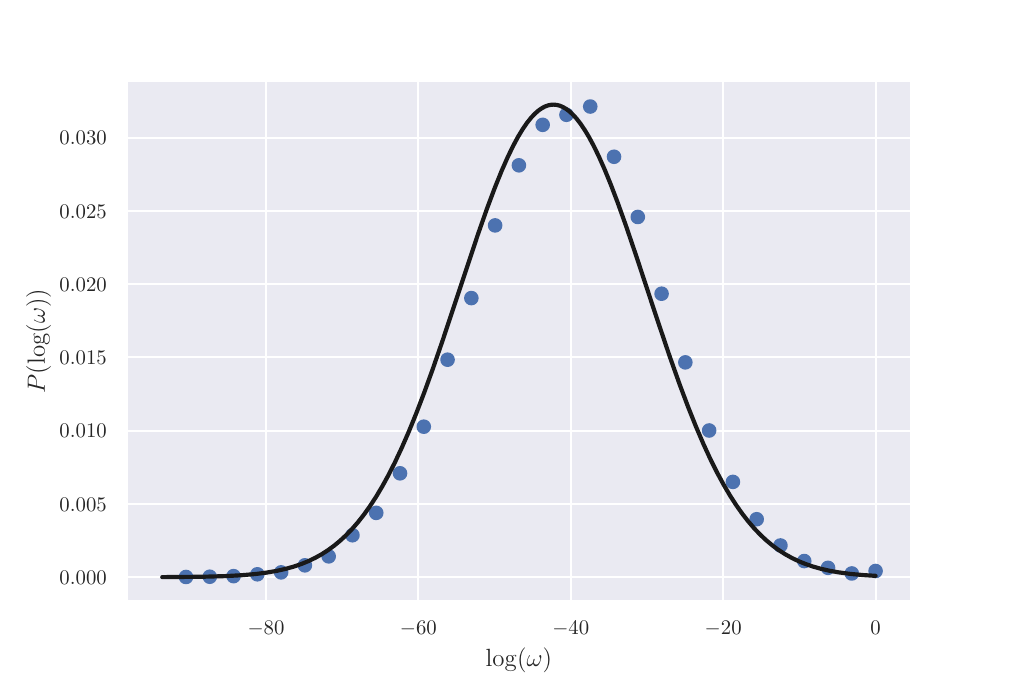}
        \caption{$p=-0.020$}
        \label{fig:sub1}
    \end{subfigure}%
    \begin{subfigure}[b]{0.5\textwidth}
        \includegraphics[width=\textwidth]{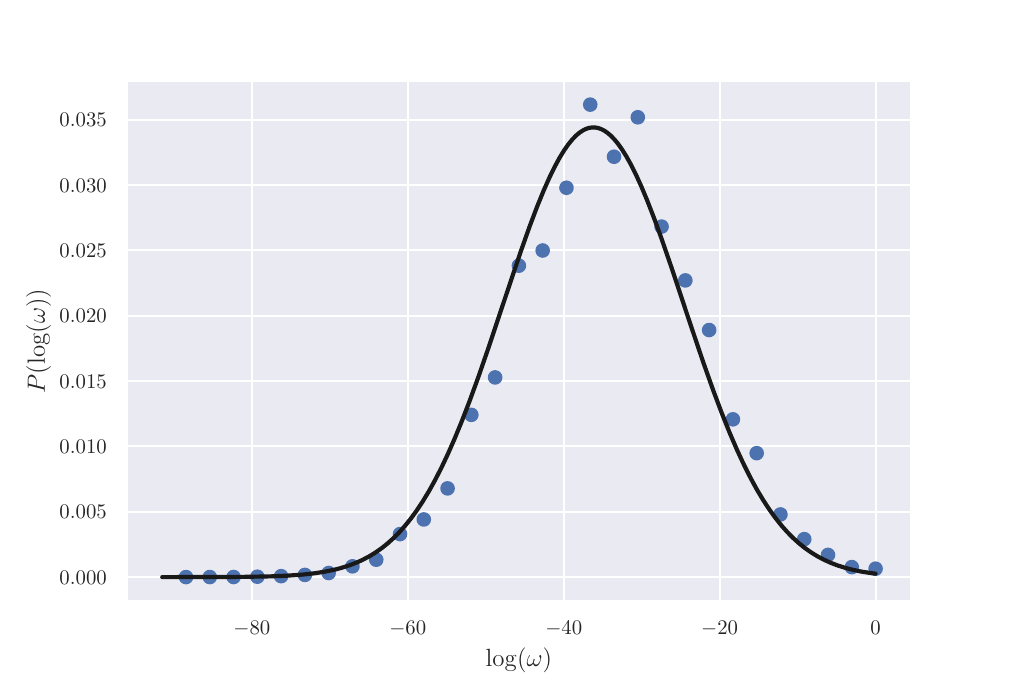}
        \caption{$p=0.0$}
        \label{fig:sub2}
    \end{subfigure}

    \begin{subfigure}[b]{0.5\textwidth}
        \centering
        \includegraphics[width=\textwidth]{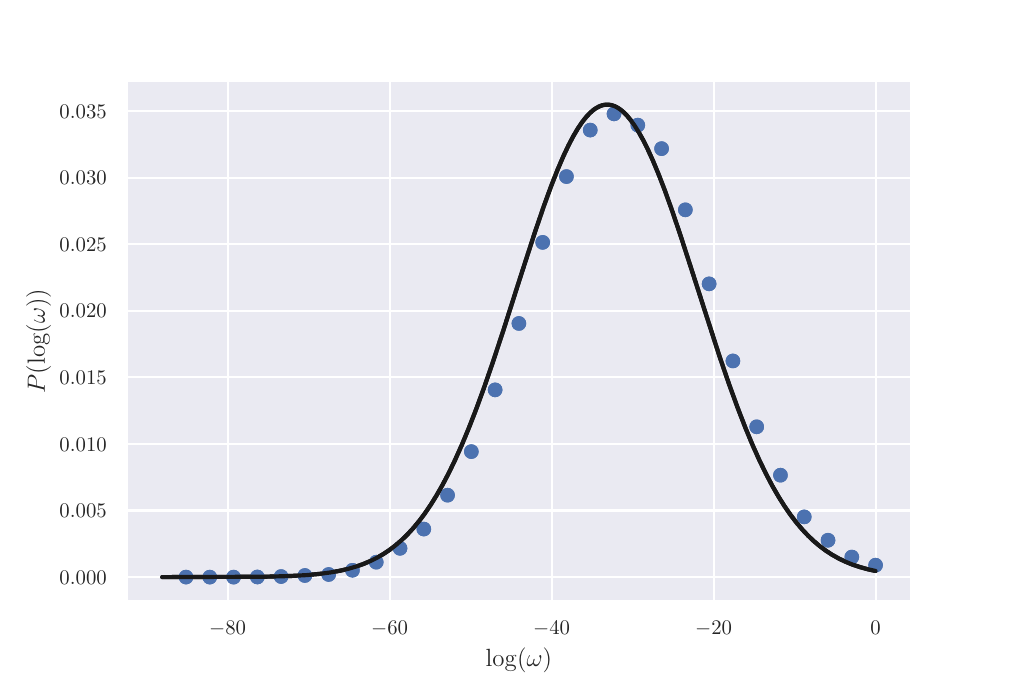}
        \caption{$p=0.020$}
        \label{fig:sub3}
    \end{subfigure}

    \caption{Relative wealth distribution for $A=0.050$ and $R=0$, for (a) $p=-0.020$, (b) $p=0.0$ and (c) $p=0.020$. $N=1000$ was used and simulations were performed until $t=10^3$ time steps.}
    \label{gauss}
\end{figure}

Beyond the analysis of specific groups, we also considered the steady state distribution of relative wealth among agents. The wealth distribution is lognormal in both progressive and regressive tax scenarios. Some wealth distributions for different tax values are shown in Figure \ref{gauss}. This result is a direct consequence of the Central Limit Theorem for the logarithm of wealth, as previously seen for $p=0$ case. The filled curves in the graphs are Gaussians with the same mean and variance as the data.

\begin{figure}
    \centering
    \begin{subfigure}[b]{0.5\textwidth}
        \includegraphics[width=\textwidth]{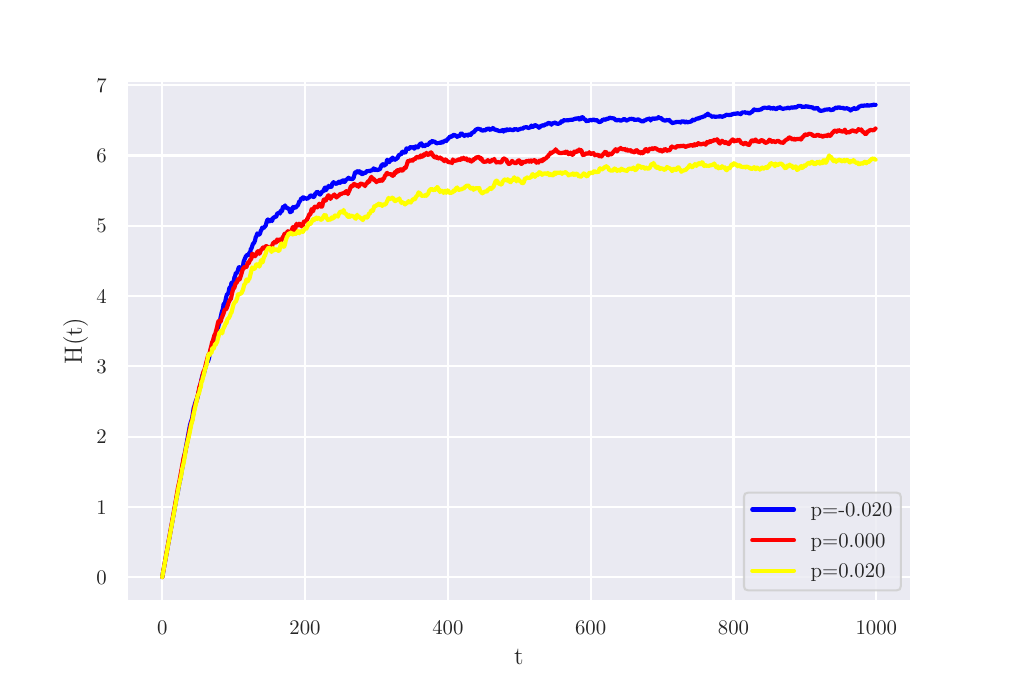}
        \caption{}
        \label{fig:sub5a}
    \end{subfigure}%
    \begin{subfigure}[b]{0.5\textwidth}
        \includegraphics[width=\textwidth]{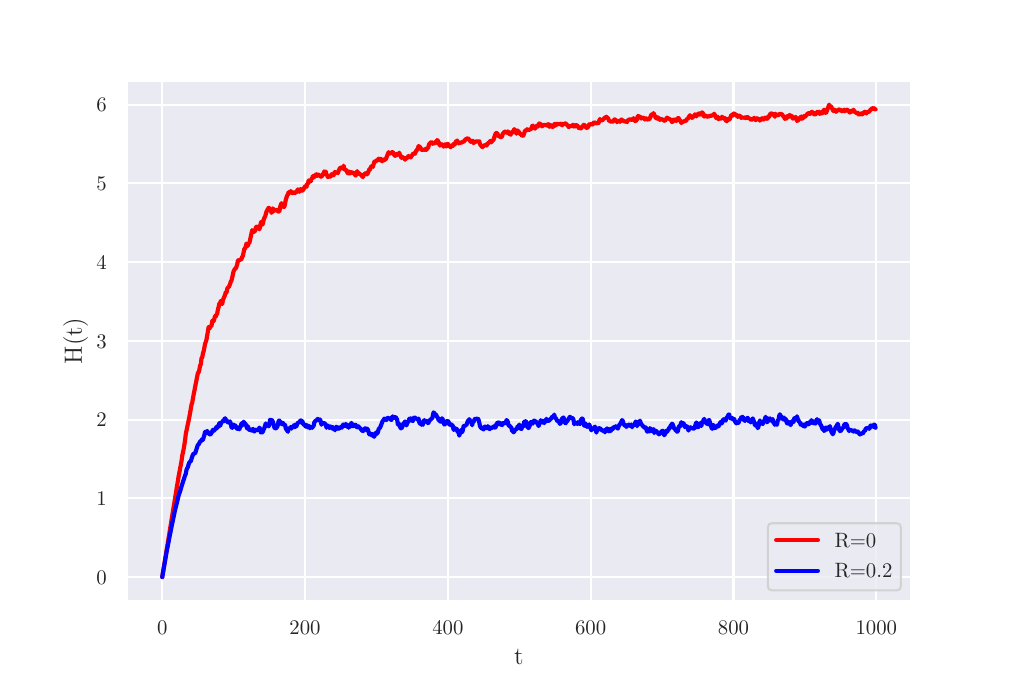}
        \caption{}
        \label{fig:sub5b}
    \end{subfigure}
    \caption{(a) Evolution in time of $H$ for $A=0.05$, $N=1000$ and $R=0$ (without redistribution). In (b) the effect of redistribution for $p=0.020$ on $H$ is evident: the wealth inequality decreases. Simulations were performed until $t=10^3$ time steps.}
    \label{ginilike}
\end{figure}

At last, we studied the degree of wealth inequality by the entropy approach \cite{Kapur}

\begin{equation}
H=ln N + \sum_{i=1}^{N} \omega_i \, ln \omega_i,
\end{equation}
where $N$ is the number of agents. This function $H \geq 0$ (non-negative) is minimized ($H=0$) for $\omega_i=1/N$ $\forall i$, the state of absolute equality. In turn, $H$ reaches a maximum ($H=lnN$) for $\omega_1=1$ and $\omega_i=0, \, i\geq 2$, a collapsed state in which total wealth is in the hands of a single agent. Outside the region of general impoverishment, our simulations reveal that $H=lnN$ (its maximum) for all values $p<0$ (regressive taxation). In contrast, lower values, dependent on $p$, are found for progressive taxes ($p>0$) even in the absence of redistribution ($R=0$). (see Figure \ref{ginilike}.)

\subsection{The roles of distributional policies}
In this subsection we shall focus on how non-uniform redistributions affects wealth dynamics. As early demonstrated by Castro de Oliveira \cite{Oliveira}, uniform redistribution plays the role of an external field conjugated to the order parameter characterizing the transition from a collapsed to an active economy. So, a $R \neq 0$ value forbids that total wealth becomes owned by a single agent, even for regressive ($p<0$) taxation.

\begin{figure}
    \centering
    \includegraphics[width=0.5\textwidth]{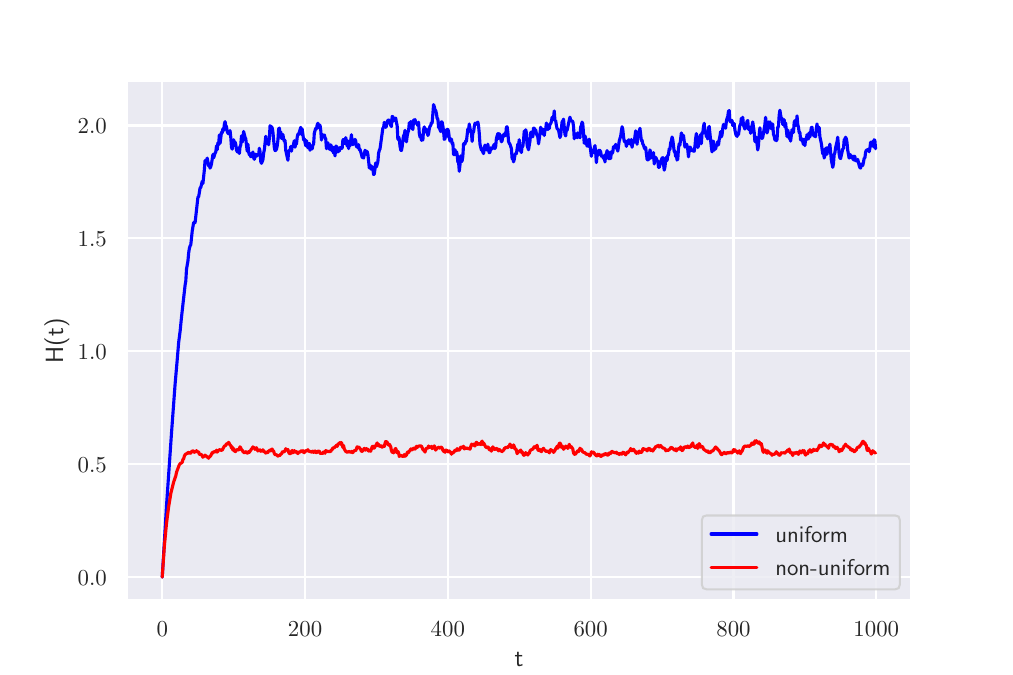}
	\caption{Evolution in time of the function H under fixed taxes and redistribution rate $R=0.20$. Uniform and progressive non-uniform (Hill type) redistributions, blue and red respectively, were considered. $N=1000$, $A=0.05$, $p=0.02$, $c=30$ were used and simulations performed until $t=10^3$ time steps.}
	\label{redistribui}
\end{figure}

We simulated a redistribution policy that prioritizes the poorer agents and compare their results with those for a uniform strategy. Accordingly, instead a constant value $R$, a non linear, sigmoidal  (Hill type) redistribution rate $R n^2/(c+n^2)$ was implemented. Here, $n$ is the agent's rank in the descending order of wealths, $c$ is a constant shaping the sigmoid and $R$ is the maximal wealth dependent redistribution rate. In Figure \ref{redistribui} are shown  evolutions in time of the $H$ function for uniform and non-uniform redistributions with $R=0.20$ fixed. As one can see, the stationary value of $H(t)$ is reached much earlier and is significantly lower under Hill redistribution in comparison with an uniform redistributive regime. Moreover, as expected, greater the redistribution $R$, smaller is the wealth concentrated in the richest agent's hands.

\subsection{Effects of tax evasion}

As a rule, the maximization of fiscal revenue has always been the major administrative goal of every government. So, income tax evasion seriously threatens this imperative, but also hinders distributional policies aimed at inequality mitigation. In contrast to taxpayers in the bottom $99\%$ of the income distribution, the top $1\%$ has at your disposal a large set of sophisticated and complex forms of tax evasion. Here, we simply assume that agents owning relative wealth $w_i \geq w_{ref}$, where $w_{ref}$ is a new constant parameter model, pay lower taxes. So, those agents are subjected to an effective tax rate $(1-\alpha)p$, with $0 \leq \alpha \leq 1$.

\begin{figure}
    \centering
    \begin{subfigure}[b]{0.5\textwidth}
        \includegraphics[width=\textwidth]{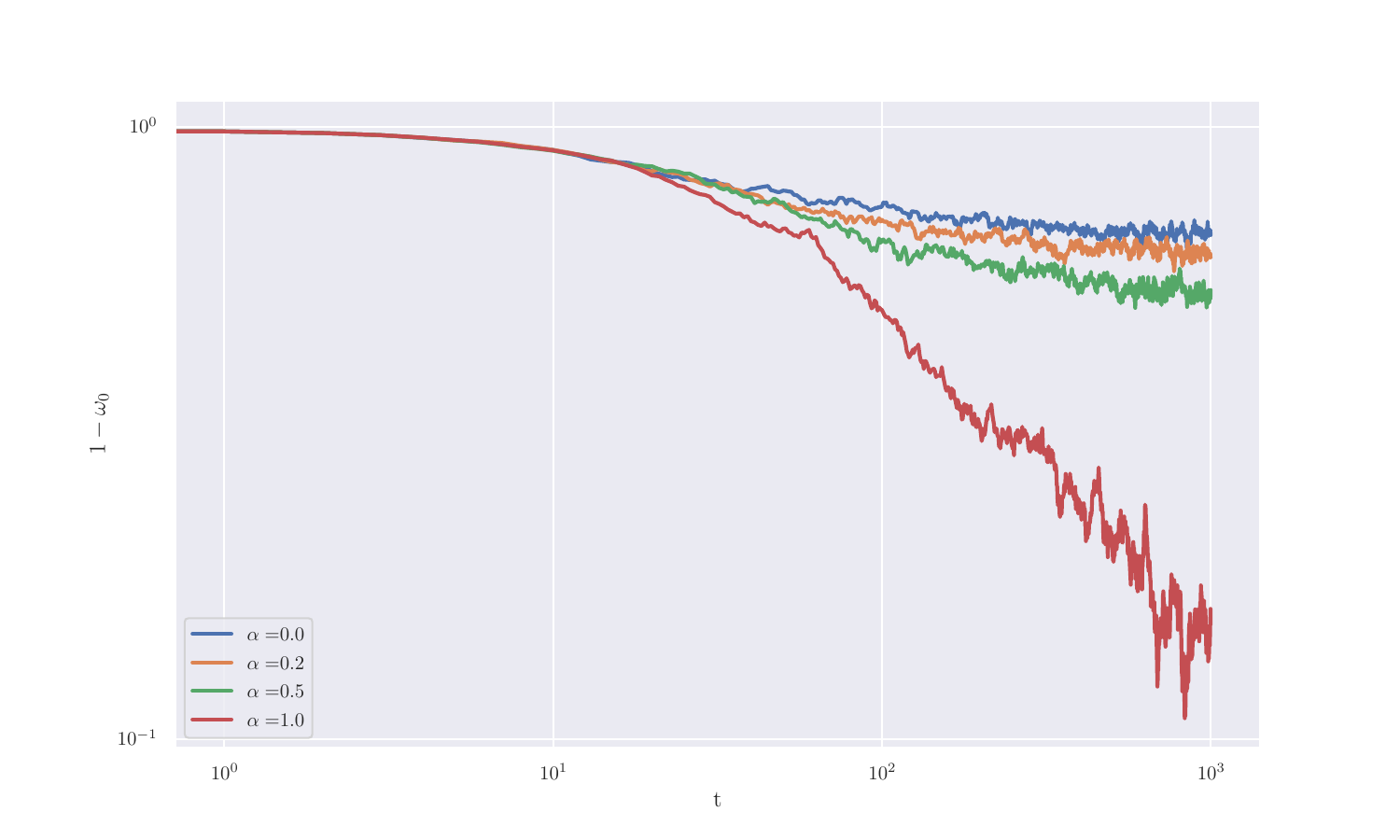}
        \caption{}
        \label{fig:sub1}
    \end{subfigure}%
    \begin{subfigure}[b]{0.5\textwidth}
        \includegraphics[width=\textwidth]{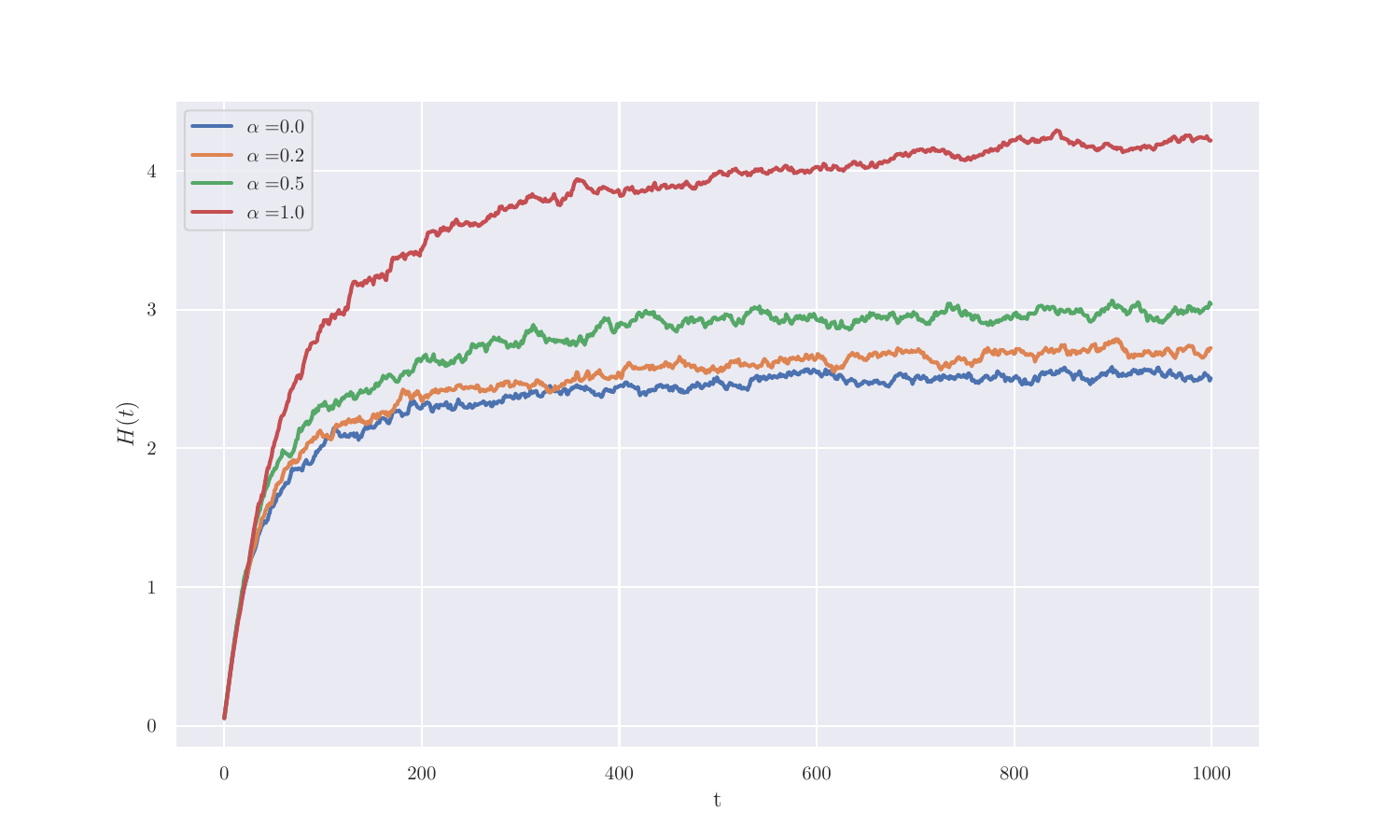}
        \caption{}
        \label{fig:sub2}
    \end{subfigure}

    \begin{subfigure}[b]{0.5\textwidth}
        \includegraphics[width=\textwidth]{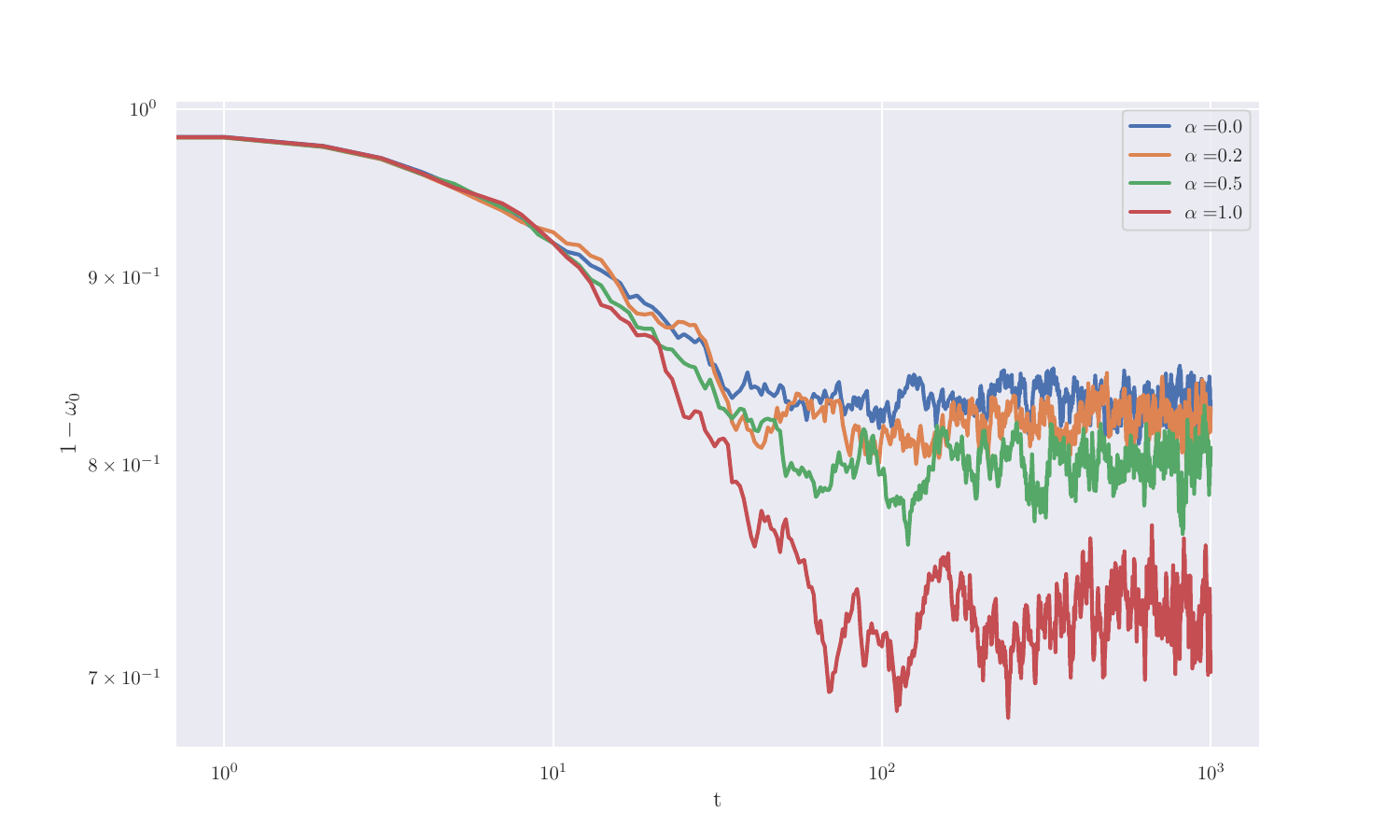}
        \caption{}
        \label{fig:sub3}
    \end{subfigure}%
    \begin{subfigure}[b]{0.5\textwidth}
        \includegraphics[width=\textwidth]{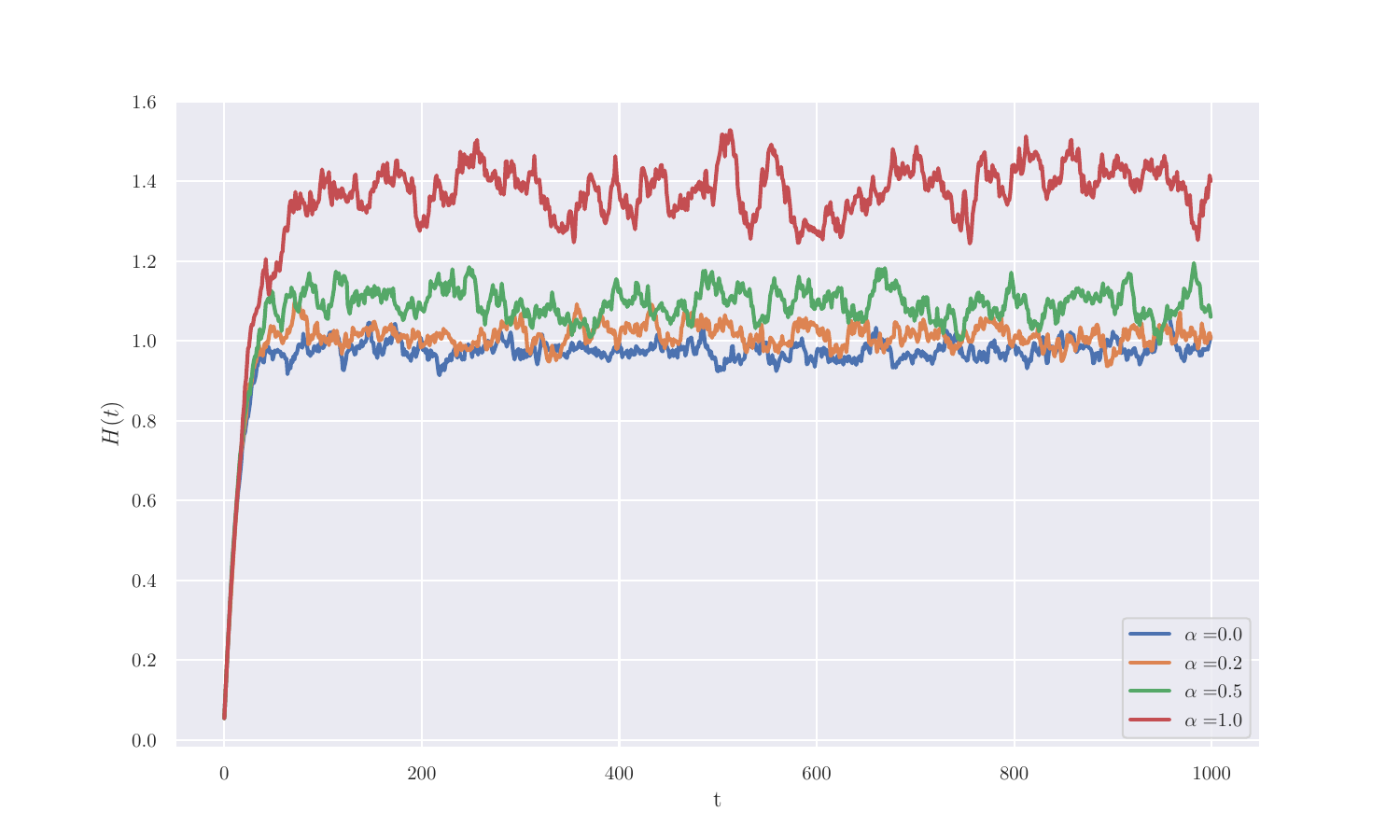}
        \caption{}
        \label{fig:sub4}
    \end{subfigure}

    \caption{Deconcentrated wealth $1-w_{max}$ and H function under regimes of fiscal evasion $\alpha \neq 0$ and a progressive tax rate $p=0.200$. (a) $1-w_{max}$, (b) $H(t)$, both with $R=0$ (redistribution absent). (c) $1-w_{max}$, (d) $H(t)$ for $R=0.2$ (uniform redistribution in action). $N=100$, $A=0.05$ and $w_{ref}=0.1$ were fixed.}
    \label{evasion}
\end{figure}

\begin{figure}
    \centering
    \begin{subfigure}[b]{0.5\textwidth}
        \includegraphics[width=\textwidth]{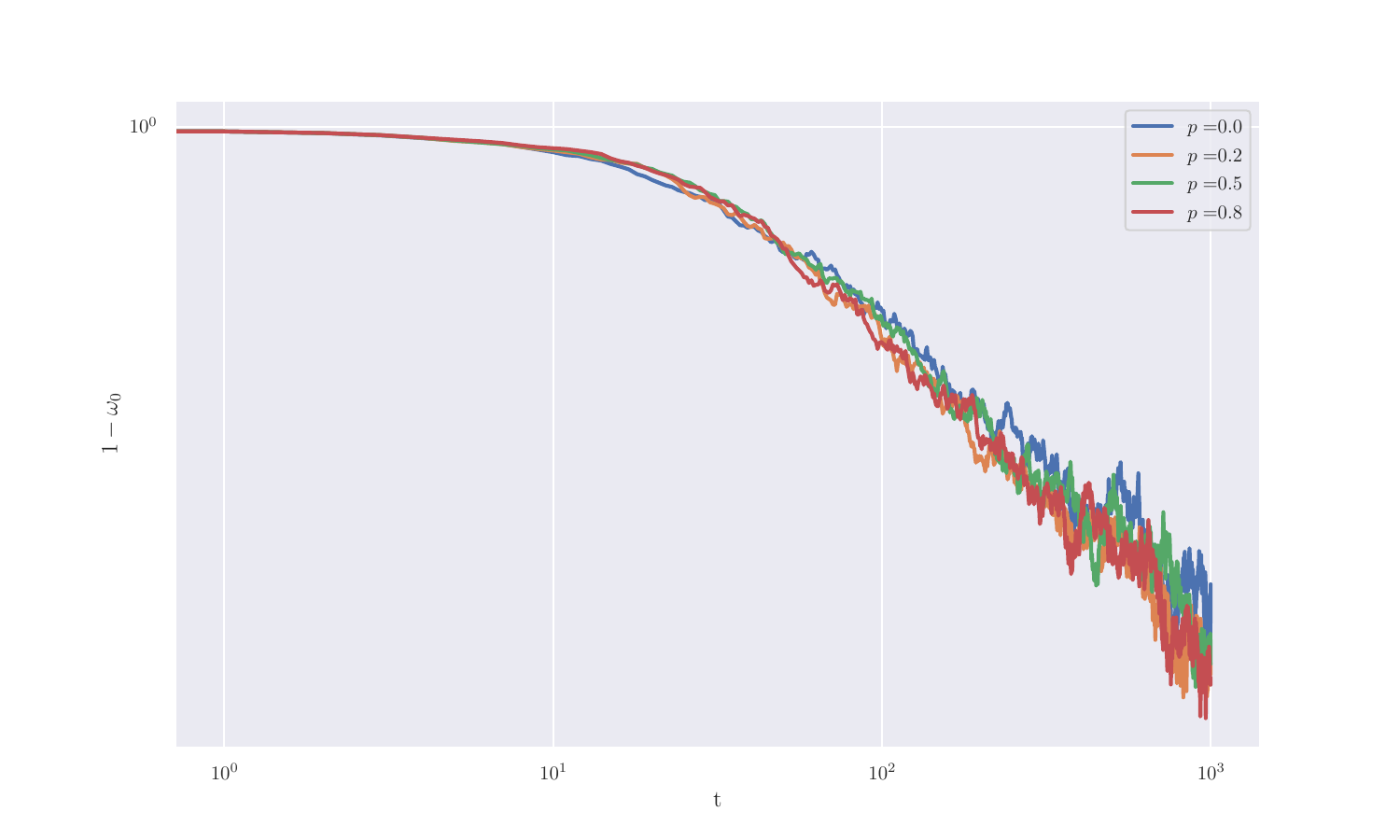}
        \caption{}
        \label{fig:sub1}
    \end{subfigure}%
    \begin{subfigure}[b]{0.5\textwidth}
        \includegraphics[width=\textwidth]{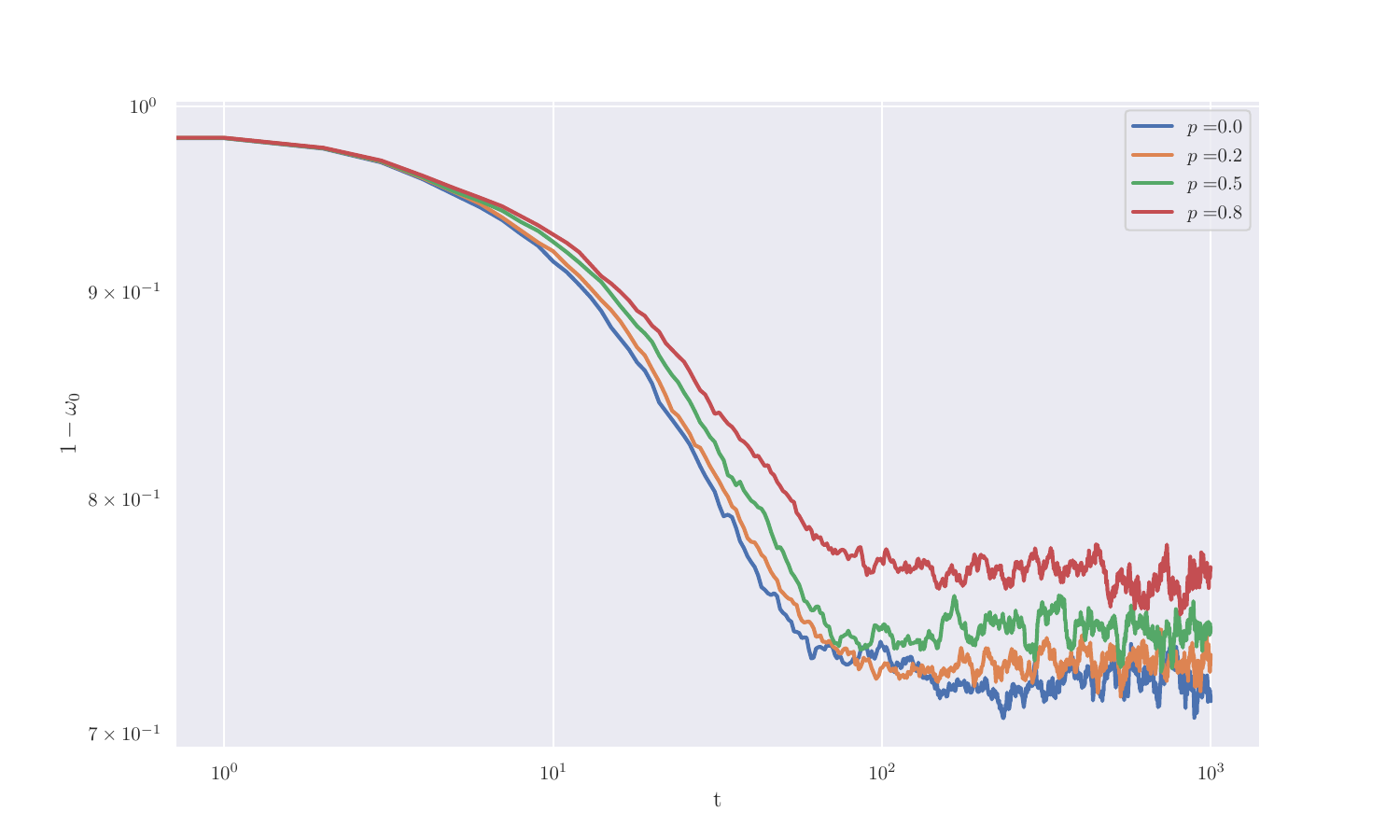}
        \caption{}
        \label{fig:sub2}
    \end{subfigure}

    \caption{Time evolution of $1-w_{max}$ under regimes of complete fiscal evasion $\alpha = 1$ and progressive tax rates. In (a) the absence of redistribution makes the system collapse to the critical curve $p=0$. (b) Redistribuition ($R \ne 0$) prevents the collapse. $N=100$, $A=0.05$, $R=0.2$ and $w_{ref}=0.1$ were fixed.}
    \label{collapse}
\end{figure}

Figure \ref{evasion} shows the evolution in time of $1-w_{max}$ for distinct values of the parameter $\alpha$ and progressive taxation ($p>0$). As expected, wealth concentration decreases with increasing tax rates but, in the absence of redistribution ($R=0$), a value $\alpha=1$ generates the economy collapse  independently of $p$ (see Figure \ref{collapse}). Thus, if at least one agent overcomes the barrier $w_{ref}$ and becomes tax free because $\alpha=1$, the total wealth concentrates only in his hands. Furthermore, complete fiscal evasion $\alpha=1$ always lead to a frozen economy at long term unless $R \neq 0$. Although redistribution prevents the collapse, our simulations demonstrate that wealth inequality strongly increases in comparison with $\alpha=0$ regimes, regardless the progressive tax rate $p$, as can be seen in Figure \ref{evasion}.

\section{Discussion}
In modern economies, tax systems play a dual pivotal role, namely, to provide primary revenue source for governments sustain their operation and to enable wealth redistribution without directly increasing the price of work. Given its centrality for economic development, provision of public goods, and achievement of equity goals, taxation stands as a very powerful instrument for social transformation, hence its appropriation by elites. Such a capture distorts the tax system and introduce mechanisms not based on and confirmed by empirical research that undermine the social pact. Therefore, understanding even simple quantitative models for wealth dynamics under taxation is imperative for us.

In this work, the effects of taxation, income redistribution, and fiscal evasion on wealth distribution were simulated through an extended version of the agent based model proposed by Castro de Oliveira \cite{Oliveira}. The model assumes that, at each discrete time step, the wealth of every agent initially grows accordingly a simple binary multiplicative process, then taxes are collected by government and a fraction of the total public revenue is uniformly redistributed among agents. In our extended version, non-uniform redistribution was considered and the possibility of fiscal evasion by richest agents allowed. Our major findings are the following.

This model economy without redistribution ($R=0$) exhibits three dynamical behaviors depending on the values of tax rates $A$ and $p$. See figures \ref{evolveW} and \ref{parts}. In the first one, called general impoverishment, all agents becomes progressively poorer. It occurs inside the region of parameter space where the trivial fixed point of the mapping (\ref{map}) is stable. This region corresponds to large total taxes ($A+p$ and $p \geq 0$) payed by agents. So, ``draconian'' taxes kill the goose that lays the golden eggs and immediately supports the liberal wail associated to the Laffer curve \cite{Laffer}. The second behavior, named enrichment with dispossession, is characterized by the rich are ever richer and the poor ever poorer. For $R=0$ this behavior is controled by the fixed point ($w_1=1, w_n=0$ $\forall n\geq 2$) of the mapping (\ref{map}). The introduction of distributive policies, i. e. $R \neq 0$ drives system outwards this collapsed economy. Finally, in the third behavior, global prosperity, the wealth $W_n$ owned by all agents increases continuously, but at different rates. We must emphasize here that a monotonic increasing of $W_n$ does not means the same for wealth share $\omega_n$. Hence, the general prosperity is also controlled by the fixed point ($w_1=1, w_n=0$ $\forall n\geq 2$) for $R=0$. Furthermore, the regime of general prosperity draws our attention to a fundamental question: wealth and inequality are not mutually excluding. Indeed, our results reveal that is possible to observe a generalized increasing of incomes ($W$), thus eradicating poverty, but with raising inequality up to the asymptotic limit in which total wealth share ($\omega$) condenses in the hands of a single agent. It is worth to mention that liberal economists tend to support the argument according to which only economic growth, not the redistribution between capital and labor, can truly higher the standards of living. Poverty, not inequality, is the problem. Even worse, the neoliberal argument claims that world poverty and inequality have both dropped. However, the empirical basis of such neoliberal argument is for a long time questioned \cite{Wade}.

Concerning the steady-state wealth distribution, it is a lognormal distribution, as shown in Figure \ref{gauss}. As demonstrated in the analysis of the $p=0$ case, the multiplicative stochastic nature of the model produces this type of distribution. Numerical results showed that this outcome is robust and prevails for both progressive and regressive taxes. The main difference is that their means decrease while their variances increase as $p>0$ (progressive tax) and/or $R \ne 0$ increases, thus flattening the distributions and expanding the "middle class" range, hence improving equality. It is already well known that a lognormal distribution fits the middle-to-bottom class in income distribution, but the model, as proposed, also shows the top of the distribution following these results, making it unable to explain the power-law tails observed in the economy. Further investigations need to be conducted to understand which missing mechanism in the model is responsible for this change in the distribution.

Once prices have both allocative and distributive functions, in market economies where substitution between capital and labor is flexible and efficient , income redistribution through tax instead wage increase, thereby labor price raising, seems to be better. The proposal of a non-uniform redistribution not only prevents evolution towards the absorbing state $\omega_{max} \rightarrow 1$ but also promotes a less unequal distribution of wealth and faster economic stationarity compared to its uniform counterpart. See Figure \ref{redistribui}. Ensuring that the most needy agents receive a greater amount than the most affluent is key to efficient redistribution. As expected, our simulations indicate that the higher is the redistribution rate $R$, lower is inequality measured by the increase in relative wealth $1 - \omega_{max}$ which is not owned by the richest agent. This indicates that not only the form of redistribution, but also the amount distributed to agents, is important to achieve a less unequal system. 

At last, it was studied a problem that transcends cultural and political boundaries: tax evasion. This phenomenon severely undermines public revenue, economic development largely financed by government’s budget, and broader social goals derived from fiscal policies. Our simulations demonstrate that financial engineering mechanisms designed for  tax evasion significantly affect wealth redistribution, as shown in Figure \ref{evasion}. On the contrary to redistribution that prevents economy from collapsing into the hands of a single agent, tax evasion reintroduces this collapse even under progressive taxation. Indeed, in the absence of distributive policies ($R=0$), if some of the richer agents no longer pay taxes ($\alpha = 1$), all curves $1-\omega_{max}$ collapse into one analogue to the Castro de Oliveira critical curve at $p=0$ and $R=0$. Thus, a totally concentrated economy emerges after a transient period that increases as a power-law. Furthermore, redistribution ($R \neq 0$) prevents such absolute concentration and the curves for $\alpha < 1$, meaning always partial evasion, have stationary values dispersed over ranges that decrease as $\alpha$ increases. So, our simulations support that the tight control of public revenue collection is of utmost importance because tax evasion strongly reduces the effects of redistribution and results in unavoidable economic inequality.

Summarizing, our results indicate that the major problem of economic inequality can not be efficiently mitigated without the concerted use of diverse fiscal policies: progressive taxation, non uniform distributive actions favoring the poorer, and strong commitment in avoid fiscal evasion able by itself to completely undermine taxation social goals. Although self-evident for many people, these results are fiercely contested by neoliberal ideologies exhibiting distinct levels of refinements but not based on `positive science'.

\section{Material and Methods}
\label{sec:methods}

\section*{Author Contributions}
Conceptualization, M.L.M.; Formal Analysis, I.N.B. and M.L.M.; Methodology, M.L.M; Software, I.N.B.; Investigation, I.N.B. and M.L.M.; Data curation, I.N.B; Writing–original draft, I.N.B.; Writing–review \&
editing, I.N.B. and M.L.M.; Project administration, M.L.M.. 

\section*{Data availability}
All relevant data are available from the corresponding author upon request.

\section*{Acknowledgments}
This work was partially supported by the Brazilian Agencies CAPES (Barros graduate fellowship), CNPq (306024/2013-6 and 400412/2014-4), and FAPEMIG (APQ-04232-10 and APQ-02710-14). This work is dedicated to the memory of Paulo Murilo Castro de Oliveira, our dear friend and master.




\end{document}